\def\rfr#1{eq. (\ref{#1})}

\def\Rfr#1{Eq. (\ref{#1})}

\def\dert#1#2{\frac{{{d}}{#1}}{{{d}}{#2}}}              

\def\bb{\bibitem}
\def\eqI{\begin{equation}}
\def\eqF{\end{equation}}
\def\eqIa{\begin{eqnarray}}
\def\eqFa{\end{eqnarray}}
\def\rp#1#2{{#1\over#2}} \def\lb#1{\label{#1}}
\def\virg#1{``#1''}

\def\bds#1{\boldsymbol{#1}}

\documentclass[11pt]{article}
\usepackage{amsmath,amsthm,amscd,amssymb}
\usepackage{latexsym}
\usepackage{graphicx,epsfig}

\begin{document}

\noindent{\bf \LARGE{Orbital effects of Sun's mass loss
and the Earth's fate}}
\\
\\
\\
{L. Iorio$^{\ast}$\\
{\it $^{\ast}$INFN-Sezione di Pisa. Address for correspondence: Viale Unit$\grave{a}$ di Italia 68
70125 Bari (BA), Italy.  \\ e-mail: lorenzo.iorio@libero.it}

\vspace{4mm}

\begin{abstract}
 I calculate  the classical
 effects induced by an isotropic mass loss $\dot M/M$ of a body on the orbital motion of a test particle around it; the present analysis is also valid  for a variation $\dot G/G$ of the Newtonian constant of gravitation.
 I perturbatively obtain negative secular rates for the osculating semimajor axis $a$, the eccentricity $e$ and the mean anomaly $\mathcal{M}$, while the argument of pericenter $\omega$  does not undergo secular precession; the longitude of the ascending node $\Omega$ and the inclination $i$ remain unchanged as well. The anomalistic period is different from  the Keplerian one, being larger than it. The true orbit, instead, expands, as shown by a numerical integration of the equations of motion in Cartesian coordinates; in fact, this is in agreement with the seemingly counter-intuitive decreasing of $a$ and $e$ because they only refer to the osculating Keplerian ellipses which approximate the trajectory at each instant.
 %
%
By assuming for the Sun $\dot M/M = -9\times 10^{-14}$  yr$^{-1}$ it turns out that the Earth's perihelion position is displaced outward by 1.3 cm
along the fixed line of apsides  after each revolution. By applying my results  to the phase in which the radius of the Sun, already moved to the Red Giant Branch of the Hertzsprung-Russell Diagram, will become as large as 1.20 AU in about $1$ Myr, I find that the Earth's perihelion position on the fixed line of the apsides  will increase by  $\approx 0.22-0.25$ AU (for $\dot M/M = -2\times 10^{-7}$ yr$^{-1}$); other researchers point towards an increase of $0.37-0.63$ AU. Mercury will be destroyed already at the end of the Main Sequence, while Venus should be engulfed in the initial phase of the Red Giant Branch phase; the orbits of the outer planets will increase by $1.2-7.5$ AU.  Simultaneous long-term numerical integrations of the equations of motion of all the major bodies of the solar system, with the inclusion of a mass-loss term in the dynamical force models as well, are required to check if the mutual N-body interactions may substantially change the picture analytically outlined here, especially in the Red Giant Branch phase in which Mercury and Venus may be removed from the integration.
\end{abstract}


\section{Introduction}
I deal with the topic of determining the classical orbital effects induced by an isotropic variation $\dot M/M$ of the mass of a central body  on the motion of a test particle; my analysis is also valid  for a change $\dot G/G$ of the Newtonian constant of gravitation. This problem, although interesting in itself, is not only an academic one because of the relevance that it may have on the ultimate destiny of planetary companions in many stellar systems in which the host star experiences a mass loss, like our Sun \cite{Sch08}.
 With respect to this aspect, my analysis may be helpful  in driving future researches towards the implementation of long-term N-body simulations including the temporal change of $GM$ as well, especially over timescales covering paleoclimate changes, up to
the Red Giant Branch (RGB) phase
in which some of the inner planets should  be engulfed by the expanding Sun.
Another problem, linked to the one investigated here, which has recently received attention is the
observationally determined secular variation of the Astronomical Unit \cite{Kra04,Sta05,Nor08,Kli08}. Moreover, increasing accuracy in astrometry pointing towards microarcsecond level \cite{IAU07}, and
long-term stability in clocks \cite{Osk06} require to consider the possibility that smaller and subtler perturbations will be soon detectable in the solar system.  Also future planetary ephemerides should take into account $\dot M/M$. Other phenomena which may show connections with the problem treated here are the secular decrease of the semimajor axes of the LAGEOS satellites, amounting to 1.1 mm d$^{-1}$, \cite{Ruby} and the increase of the lunar orbit's eccentricity \cite{luna}. However, a detailed analysis of all such issues is beyond the scope of this paper.

Many treatments of the mass loss-driven  orbital dynamics in the framework of the Newtonian mechanics, based on different approaches and laws of variation of the central body's mass, can be found in literature; see, e.g., \cite{Stro,Jea24,Jea29,Arm53,Haj63,Haj66,Khol,Dep83,Kev96,Kra04,Nor08} and references therein. 

The plan of the paper is as follows. Section \ref{minchia} is devoted to a theoretical description of the phenomenon in a two-body scenario. By working in the Newtonian framework, I will  analytically work out the changes after one orbital revolution experienced by all the Keplerian orbital elements of a test particle moving in the gravitational field of a central mass experiencing a variation of its $GM$ linear in time. Then, I will clarify the meaning of the results obtained by  performing a numerical integration of the equations of motion in order to visualize the true trajectory followed by the planet.  Concerning the method adopted, I will use the Gauss perturbation equations \cite{Bert,roy}, which are valid for generic disturbing accelerations depending on position, velocity and time,  the \virg{standard} Keplerian orbital elements (the Type I according to, e.g., \cite{Khol}) with the eccentric anomaly $E$ as \virg{fast} angular variable. Other approaches and angular variables like, e.g. the Lagrange perturbation equations \cite{Bert,roy}, the Type II orbital elements \cite{Khol} and the mean anomaly $\mathcal{M}$ could be used, but, in my opinion, at a price of major conceptual and computational difficulties\footnote{Think, e.g.,  about the cumbersome expansions in terms of the mean anomaly and the Hansen coefficients, the subtleties concerning the choice of the independent variable in the Lagrange equations for the semimajor axis and the eccentricity \cite{Bert}.}. With respect to possible connections with realistic situations, it should be noted that, after all, the Type I orbital elements are usually determined or improved in standard data reduction analyses of the motion of planets and (natural and artificial) satellites. Instead, my approach should, hopefully, appear more transparent and easy to interpret, although, at first sight, some counter-intuitive results concerning the semimajor axis and the eccentricity will be obtained; moreover, for the chosen time variation of the mass of the primary, no approximations are used in the calculations which are quite straightforward.  However, it is important to stress that such allegedly puzzling features are only seemingly paradoxical because they will turn out to be in agreement with numerical integrations of the equations of motion, as explicitly shown by the Figures depicted.  Anyway, the interested reader is advised to look also at \cite{Khol} for a different approach.
%
In Section \ref{evol} I will apply my results to the future Sun-Earth scenario and to the other planets of the solar system.
Section \ref{quattro} summarizes my results.
%

%
%
\section{Analytical calculation of the orbital effects by $\dot\mu/\mu$}\lb{minchia}

By defining
\eqI\mu\doteq GM \eqF at a given epoch $t_0$,
the acceleration of a test particle orbiting a central body experiencing a variation of $\mu$ is, to first order in $t-t_0$,
\eqI \bds{A} =-\rp{\mu(t)}{r^2}\bds{\hat r}\approx -\rp{\mu}{r^2}\left[1 + \left(\rp{\dot\mu}{\mu}\right)(t-t_0)\right]\bds{\hat r},\lb{accel}\eqF
with $\dot\mu\doteq\dot\mu|_{t=t_0}$. $\dot\mu$ will be assumed constant throughout the temporal interval of interest $\Delta t = t-t_0$, as it is, e.g., the case for most of the remaining lifetime of the Sun as a Main Sequence (MS) star \cite{Sch08}.    Note that $\dot\mu$ can, in principle, be due to a variation of both the Newtonian gravitational constant $G$ and the mass $M$ of the central body, so that
\eqI \rp{\dot\mu}{\mu} = \rp{\dot G}{G} + \rp{\dot M}{M}.\eqF
Moreover, while the orbital angular momentum is conserved, this does not happen for the energy.

By limiting ourselves to realistic astronomical scenarios like our solar system, it is quite realistic to assume that
\eqI \left(\rp{\dot\mu}{\mu}\right)(t-t_0)\ll 1\lb{condiz}\eqF  over most of its remaining lifetime:
 indeed, since  $\dot M/M$ is of the order of\footnote{About $80\%$ of such a mass-loss is due to the core nuclear burning, while the remaining $20\%$ is due to average solar wind.} $10^{-14}$ yr$^{-1}$ for the Sun \cite{Sch08}, the condition of \rfr{condiz} is satisfied for the remaining\footnote{The age of the present-day MS Sun is 4.58 Gyr, counted from its zero-age MS star model \cite{Sch08}.} $\approx$ 7.58 Gyr before the Sun will approach the  RGB tip in the Hertzsprung-Russell Diagram (HRD). Thus, I can treat it perturbatively with the standard methods of celestial mechanics.

 The unperturbed  Keplerian ellipse at epoch $t_0$, assumed coinciding with the time of the passage at perihelion $t_p$, is characterized by
\begin{equation}
\begin{array}{lll}

r = a(1-e\cos E),\\\\
dt = \left(\rp{1-e\cos E}{n}\right)dE,\\\\
\cos f = \rp{\cos E - e}{1-e\cos E},\\\\
\sin f = \rp{\sqrt{1-e^2}\sin E}{1-e\cos E},
\end{array}\lb{cofi}
 \end{equation}
where $a$ and $e$ are the semimajor axis and the eccentricity, respectively, which fix the size and the shape of the unchanging Keplerian orbit, $n=\sqrt{\mu/a^3}$ is its unperturbed Keplerian mean motion, $f$ is the true anomaly, reckoned from the pericentre, and $E$ is the eccentric anomaly.
 \Rfr{cofi} characterizes the path followed by the particle for any $t>t_p$ if the mass loss would suddenly cease at $t_p$. Instead, the true path will be, in general, different from a closed ellipse because of the perturbation induced by $\dot\mu$ and the orbital parameters of the osculating ellipses approximating the real trajectory at each instant of time will slowly change in time.

\subsection{The semimajor axis and the eccentricity}
The Gauss equation for the variation of the semimajor axis $a$ is   \cite{Bert,roy}
\eqI\dert a t = \rp{2}{n\sqrt{1-e^2}} \left[e A_r\sin f +A_{\tau}\left(\rp{p}{r}\right)\right],\lb{gaus_a}\eqF
where $A_{r}$  and $A_{\tau}$ are the radial and transverse, i.e. orthogonal to the direction of $\bds{\hat r}$, components, respectively, of the disturbing acceleration, and $p\doteq a(1-e^2)$ is the semilatus rectum.
In the present case
\eqI A= A_r = -\rp{\dot\mu}{r^2}(t-t_p),\lb{dist}\eqF i.e. there is an entirely radial perturbing acceleration. For $\dot\mu<0,$ i.e. a decrease in the body's $GM$, the total gravitational attraction felt by the test particle, given by \rfr{accel}, is reduced with respect to the epoch $t_p$.
In order to have the rate of the semimajor axis averaged over one (Keplerian) orbital revolution \rfr{dist} must be inserted into \rfr{gaus_a},   evaluated onto the unperturbed Keplerian ellipse with \rfr{cofi} and finally integrated over $ndt/2\pi$ from 0 to $2\pi$ because $n/2\pi\doteq 1/P^{\rm Kep}$ (see below).
Note that, from \rfr{cofi}, it can be obtained
\eqI t-t_p = \rp{E -e\sin E}{n}.\eqF
As a result, I have\footnote{Recall that the integration is taken over the unperturbed Keplerian ellipse: that is why a and e are kept out of the integral in \rfr{arate} and in the following averages.}
\eqI\left\langle\dert a t\right\rangle  =  -\rp{e}{\pi}\left(\rp{\dot\mu}{\mu}\right)a
\int_0^{2\pi} \rp{\left(E -e\sin E\right)\sin E}{\left(1-e\cos E\right)^2} dE = 2\left(\rp{e}{1-e}\right)\left(\rp{\dot\mu}{\mu}\right)a.\lb{arate}
\eqF
Note that if $\mu$ decreases $a$ gets reduced as well: $\left\langle\dot a\right\rangle< 0$. This may be seemingly bizarre and counter-intuitive, but, as it will be shown  later, it is not in contrast with the true orbital motion.

The Gauss equation for the variation of the eccentricity is  \cite{Bert,roy}
\eqI \dert e t  = \rp{\sqrt{1-e^2}}{na}\left\{A_r\sin f + A_{\tau}\left[\cos f + \rp{1}{e}\left(1 - \rp{r}{a}\right)\right]\right\}.\lb{gaus_e}\eqF
For $A=A_r$, it reduces to
\eqI\dert e t = \left(\rp{1-e^2}{2ae}\right)\dert a t,\eqF
so that
\eqI\left\langle\dert e t\right\rangle = (1+e)\left(\rp{\dot\mu}{\mu}\right);\lb{erate}\eqF
also the eccentricity gets smaller for $\dot\mu<0$.

 As a consequence of the found variations of the osculating semimajor axis and the eccentricity, the osculating orbital angular momentum per unit mass, defined by $L^2 \doteq \mu a(1-e^2)$, remains constant: indeed, by using \rfr{arate} and \rfr{erate}, it turns out
\eqI \left\langle\rp{d L^2}{dt}\right\rangle = \mu\left\langle\dot a\right\rangle(1-e^2)-2\mu a e \left\langle\dot e\right\rangle = 0.\lb{momang}\eqF

The osculating total energy $\mathcal{E}\doteq-\mu/2a$ decreases according to
\eqI\left\langle\dert {\mathcal{E}} t\right\rangle = \rp{\mu}{2 a^2}\left\langle\dot a\right\rangle = \left(\rp{e}{1-e}\right)\rp{\dot\mu}{a}.\lb{enosc}\eqF

Moreover, the osculating Keplerian period \eqI P^{\rm Kep}\doteq 2\pi\sqrt{\rp{a^3}{\mu}},\eqF which, by definition, yields the time elapsed between two consecutive perihelion crossings in absence of perturbation, i.e. it is the time required to describe a fixed osculating Keplerian ellipse, decreases according to
\eqI\left\langle\dert {P^{\rm Kep}} t\right\rangle = \rp{3}{2}P^{\rm Kep}\rp{\left\langle\dot a\right\rangle}{a}=\rp{6\pi e\dot\mu}{(1-e)}\left(\rp{a}{\mu}\right)^{3/2}.\lb{pke}\eqF As I will show, also such a result is not in contrast with the genuine orbital evolution.

\subsection{The pericentre, the node and the inclination}
The Gauss equation for the variation of the pericentre $\omega$ is  \cite{Bert,roy}
\eqI
\dert\omega t  = \rp{\sqrt{1-e^2}}{nae}\left[-A_r\cos f + A_{\tau}\left(1+\rp{r}{p}\right)\sin f\right]-\cos i\dert\Omega t,\lb{gaus_o}
\eqF
where $i$ and $\Omega$ are the the inclination and the longitude of the ascending node, respectively, which fix the orientation of the osculating ellipse in the inertial space.
Since $d\Omega/dt$ and $di/dt$ depend on the normal component $A_{\nu}$ of the disturbing acceleration, which is absent in the present case, and $A=A_r$, I have
\eqI
\left\langle\dert \omega t\right\rangle =\rp{\sqrt{1-e^2}}{2\pi e}\left(\rp{\dot\mu}{\mu}\right)\int_0^{2\pi}\rp{(E-e\sin E)(\cos E- e)}{(1-e\cos E)^2}dE=0:
\eqF
the osculating ellipse does not change its orientation in the orbital plane, which, incidentally, remains fixed in the inertial space because $A_{\nu}=0$ and, thus, $d\Omega/dt = di/dt = 0$.

\subsection{The mean anomaly}
The Gauss equation for the mean anomaly $\mathcal{M}$, defined as $\mathcal{M} \doteq n(t-t_p)$,   \cite{Bert,roy}
is
\eqI\dert {\mathcal{M}} t = n - \rp{2}{na} A_r\rp{r}{a} -\sqrt{1-e^2}\left(\dert\omega t + \cos i \dert\Omega t\right).\lb{gaus_M}\eqF
It turns out that, since
\eqI -\rp{2}{na}A_r\rp{r}{a}dt = \rp{2\dot\mu}{n^3 a^3}(E-e\sin E)dE,\eqF
then
\eqI \left\langle\dert {\mathcal{M}} t\right\rangle = n + 2\pi\left(\rp{\dot\mu}{\mu}\right);\eqF the mean anomaly changes uniformly in time at a slower rate with respect to the unperturbed Keplerian case for $\dot\mu< 0$.
\subsection{Numerical integration of the equations of motion and explanation of the seeming contradiction with the analytical results}
At first sight, the results obtained here may be rather confusing: if the gravitational attraction of the Sun reduces in time because of its mass loss the orbits of the planets should expand (see the trajectory plotted in Figure \ref{picture}, numerically integrated with MATHEMATICA), while I obtained that the semimajor axis and the eccentricity undergo secular decrements.
\begin{figure}
\includegraphics[width=\columnwidth]{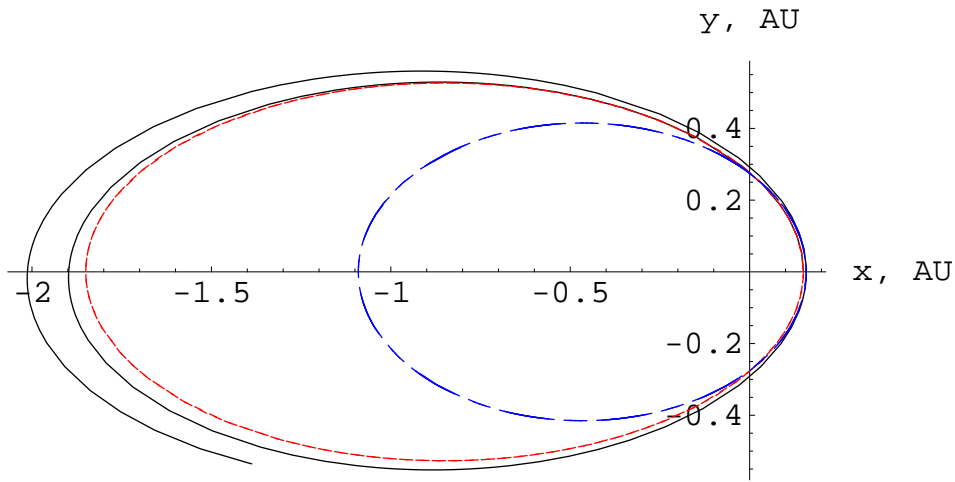}
 \caption{\label{picture} Black continuous line: true trajectory obtained by numerically integrating with MATHEMATICA the perturbed equations of motion in Cartesian  coordinates over 2 yr; the disturbing acceleration of \rfr{accel} has been adopted. The planet starts from the perihelion on the \textsc{x} axis. Just for illustrative purposes, a mass loss rate of the order of $10^{-2}$ yr$^{-1}$ has been adopted for the Sun; for the planet initial conditions corresponding to $a=1$ AU, $e=0.8$ have been chosen. Red dashed line: unperturbed Keplerian ellipse at $t = t_0= t_p$. Blue dash-dotted line: osculating Keplerian ellipse after the first perihelion passage. As can be noted, its semimajor axis and eccentricity are clearly smaller than those of the initial unperturbed ellipse. Note also that after 2 yr  the planet has not yet reached the perihelion as it would have done in absence of mass loss, i.e. the true orbital period is longer than the Keplerian one of the osculating red ellipse.}
\end{figure}
Moreover, I found that the Keplerian period $P^{\rm Kep}$
decreases,  while one would expect that the orbital period increases.

In fact, there is no contradiction, and my analytical results do yield us realistic information on the true evolution of the planetary motion. Indeed, $a$, $e$ and $P^{\rm Kep}$ refer to the osculating Keplerian ellipses which, at any instant, approximate the true trajectory; it, instead, is not an ellipse, not being  bounded. Let us start at $t_p$ from the osculating pericentre of the Keplerian ellipse corresponding to chosen initial conditions: let us use a heliocentric frame with the  \textsc{x} axis oriented along the osculating pericentre. After a true revolution, i.e. when the true radius vector of the planet has swept an angular interval of $2\pi$, the planet finds itself again on the  \textsc{x} axis, but at a larger distance from the starting point because of the orbit expansion induced by the Sun's mass loss. It is not difficult to understand that the osculating Keplerian ellipse approximating the trajectory at this perihelion passage is oriented as before because there is no variation of the (osculating) argument of pericentre,  but has smaller semimajor axis and eccentricity.
And so on, revolution after revolution, until the perturbation theory can be applied, i.e. until $\dot\mu/\mu (t-t_p)<<1$.  In Figure \ref{picture} the situation described so far is qualitatively illustrated. Just for illustrative purposes I enhanced the overall effect by assuming  $\dot\mu/\mu \approx 10^{-2}$ yr$^{-1}$ for the Sun; the initial conditions for the planet correspond to an unperturbed Keplerian ellipse with $a=1$ AU, $e=0.8$ with the present-day value of the Sun's mass in one of its foci. It is apparent that the initial osculating red dashed ellipse has larger $a$ and $e$ with respect to the second osculating blue dash-dotted ellipse.
Note also that the true orbital period, intended as the time elapsed between two consecutive crossings of the perihelion, is larger than the  unperturbed Keplerian one of the initial red dashed osculating ellipse, which would amount to 1 yr for the Earth: indeed, after 2 yr the planet has not yet reached the perihelion for its second passage.

Now, if I compute the radial change $\Delta r(E)$ in the osculating radius vector as a function of the eccentric anomaly $E$ I can gain useful insights concerning how much the true path has expanded after two consecutive perihelion passages.
From the Keplerian expression of the Sun-planet distance
\eqI r = a(1-e\cos E)\eqF one gets the radial component of the orbital perturbation  expressed in terms of the eccentric anomaly $E$
\eqI \Delta r (E) = (1-e\cos E)\ \Delta a-a\cos E\ \Delta e + ae\sin E\ \Delta E;\eqF
it agrees with the results obtained in, e.g., \cite{Cas93}.
Since
\eqI
\left\{
\begin{array}{lll}
\Delta a & = & -\rp{2ae}{n}\left(\rp{\dot\mu}{\mu}\right)\left(\rp{\sin E - E\cos E}{1-e\cos E}\right),\\\\
\Delta e & = & -\rp{(1-e^2)}{n}\left(\rp{\dot\mu}{\mu}\right)\left(\rp{\sin E - E\cos E}{1-e\cos E}\right),\lb{ecces}\\\\
\Delta E & = & \left(\rp{\Delta {\mathcal{M}} +\sin E\ \Delta e  }{1-e\cos E}\right)=\rp{1}{n}\left(\rp{\dot\mu}{\mu}\right)\left[\mathcal{A}(E)+\mathcal{B}(E)+\mathcal{C}(E)\right],
 \end{array}
\right.
\eqF
with
\eqI
\left\{
 \begin{array}{lll}
\mathcal{A}(E) & = & \rp{E^2 + 2e(\cos E -1)}{1-e\cos E},\\\\
\mathcal{B}(E) & = & \left(\rp{1-e^2}{e}\right)\left[\rp{1+e - (1+e)\cos E - E\sin E}{(1-e\cos E)^2}\right],\\\\
\mathcal{C}(E) & = & -\rp{(1-e^2)\sin E(\sin E - e\cos E)}{(1-e\cos E)^2},
\end{array}
\right.
\eqF
%
%
%
it follows
\eqI \Delta r(E) = \rp{a}{n}\left(\rp{\dot\mu}{\mu}\right)\left[\mathcal{D}(E)+\mathcal{F}(E)\right], \lb{mega}\eqF
with
\eqI
\left\{
\begin{array}{lll}
\mathcal{D}(E) & = & e\left[-2(\sin E-E \cos E) + \rp{\sin E\left[E^2 + 2e(\cos E -1)\right]}{1-e\cos E} -\rp{(1-e^2)\sin^2 E(\sin E-e\cos E)}{(1-e\cos E)^2} \right],\\\\
\mathcal{F}(E) & = & \left(\rp{1-e^2}{1-e\cos E}\right)\left\{
\cos E(\sin E - E \cos E) + \sin E\left[\rp{1+e - (1+e)\cos E-E \sin E}{1-e\cos E}\right]
\right\}.
\end{array}
\right.
\lb{mega2}\eqF
It turns out from \rfr{mega} and \rfr{mega2}  that, for $E>0$, $\Delta r(E)$ never vanishes; after one  orbital revolution, i.e. after that an angular interval of $2\pi$ has been swept by the (osculating) radius vector, a net increase of the radial (osculating) distance occurs according to\footnote{According to \rfr{mega} and \rfr{mega2}, $\Delta r (0)=0$.} \eqI\Delta r (2\pi) - \Delta r (0)= \Delta r (2\pi)  = -\rp{2\pi}{n}a\left(\rp{\dot\mu}{\mu}\right)(1-e).\lb{deltaerre}\eqF
This analytical result is qualitatively confirmed by the difference\footnote{Strictly speaking, $\Delta r$ and the quantity plotted in Figure \ref{deltaerref} are different objects, but, as the following discussion will clarify, I can assume that, in practice, they are the same.} $\Delta r(t)$ between the radial distances obtained from the solutions of two numerical integrations  of the equations of motion over 3 yr with and without $\dot\mu/\mu$; the initial conditions are the same. For illustrative purposes I used $a=1$ AU, $e=0.01$, $\dot\mu/\mu=-0.1$ yr$^{-1}$. The result is depicted in Figure \ref{deltaerref}.
\begin{figure}
\includegraphics[width=\columnwidth]{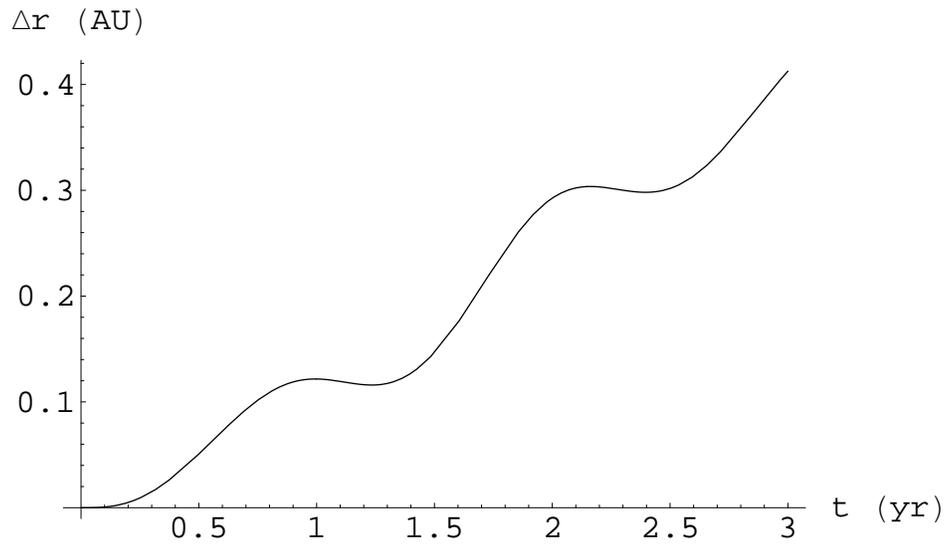}
\caption{\label{deltaerref}  Difference $\Delta r(t)$ between the radial distances obtained from the solutions of two numerical integrations  with MATHEMATICA of the equations of motion over 3 yr  with and without $\dot\mu/\mu$; the initial conditions are the same. Just for illustrative purposes a mass loss rate of the order of $-10^{-1}$ yr$^{-1}$ has been adopted for the Sun; for the planet initial conditions corresponding to $a=1$ AU, $e=0.01$ have been chosen. The cumulative increase of the Sun-planet distance induced by the mass loss is apparent.}
\end{figure}
Note also that  \rfr{mega} and \rfr{mega2} tell us that the shift at the aphelion is
\eqI\Delta r(\pi)=\rp{1}{2}\left(\rp{1+e}{1-e}\right)\Delta r(2\pi),\lb{aphe}\eqF in agreement with Figure \ref{picture} where it is 4.5 times larger than the shift at the perihelion.

Since Figure \ref{picture} tells us that the orbital period gets larger than the Keplerian one, it means that the true orbit must somehow remain behind with respect to the Keplerian one. Thus, a negative perturbation $\Delta\tau$ in the transverse direction must occur as well; see Figure \ref{AUD}.
\begin{figure}
\includegraphics[width=\columnwidth]{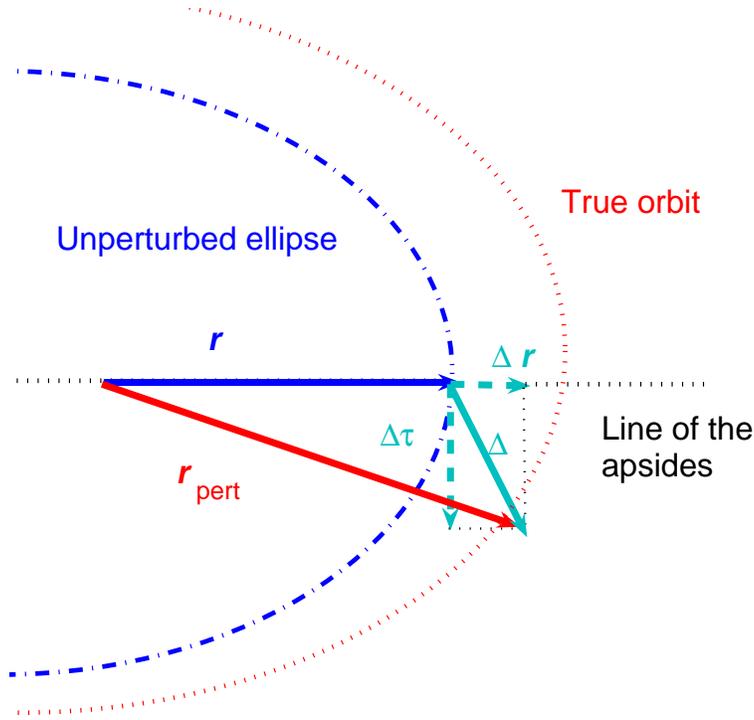}
\caption{\label{AUD}  Radial and transverse perturbations $\bds{\Delta r}$ and $\bds{\Delta \tau}$ of the Keplerian radius vector (in blue); the presence of the transverse perturbation $\bds{\Delta \tau}$  makes the real orbit (in red) lagging behind the Keplerian one.}
\end{figure}

Let us now analytically compute it.
According to \cite{Cas93}, it can be used
\eqI\Delta\tau = \rp{a\sin E}{\sqrt{1-e^2}} + a\sqrt{1-e^2}\ \Delta E + r(\Delta\omega+\Delta\Omega\cos i).\lb{caso}\eqF
By recalling that, in the present case, $\Delta\Omega=0$ and using
\eqI\Delta\omega= -\rp{\sqrt{1-e^2}}{ne}\left(\rp{\dot\mu}{\mu}\right)\left[\rp{1 + e - (1 + e)\cos E - E \sin E}{1-e\cos E}\right],\lb{peris}\eqF
it is possible to obtain from \rfr{ecces} and \rfr{peris}
\eqI\Delta\tau(E) = \rp{a}{n}\left(\rp{\dot\mu}{\mu}\right)\rp{\sqrt{1-e^2}}{\left(1-e\cos E\right)}\left[\mathcal{G}(E) + \mathcal{H}(E)+\mathcal{I}(E)+\mathcal{J}(E)+\mathcal{K}(E)\right],\lb{tra}\eqF
with
\eqI
\left\{
 \begin{array}{lll}
\mathcal{G}(E) & = & \sin E(E\cos E -\sin E),\\\\
\mathcal{H}(E) & = & \rp{(1-e\cos E)}{e}\left[(1+e)(\cos E-1) + E\sin E\right],\lb{vuuu}\\\\
\mathcal{I}(E) & = & E^2 + 2 e (\cos E-1),\\\\
\mathcal{J}(E) & = &\sin E\left[\rp{(1-e^2)(e\cos E-\sin E)}{1-e\cos E}\right]\\\\
\mathcal{K}(E) & = & \left(\rp{1-e^2}{e}\right)\left[\rp{(1+e)(1-e\cos E) - E\sin E}{1-e\cos E}\right].
\end{array}
\right.
\eqF
%
%
%
%
It turns out from \rfr{tra} and \rfr{vuuu} that, for $E>0$, $\Delta \tau(E)$ never vanishes; at the time of perihelion passage
\eqI\Delta\tau(2\pi) -\Delta\tau(0)=\rp{4\pi^2}{n}a\left(\rp{\dot\mu}{\mu}\right)\sqrt{\rp{1+e}{1-e}}<0.\eqF
This means that when the Keplerian path has reached the perihelion, the perturbed orbit is still behind it.
 Such features are qualitatively confirmed by Figure \ref{picture}.

From a vectorial point of view, the radial and transverse perturbations to the Keplerian radius vector $\bds r$ yield a correction
\eqI \bds\Delta = \Delta r\ \bds{\hat{r}} + \Delta\tau\ \bds{\hat{\tau}},\eqF so that
\eqI \bds r_{\rm pert} = \bds r + \bds\Delta.\eqF The length of $\bds\Delta$ is
\eqI\Delta(E) =\sqrt{\Delta r(E)^2 + \Delta \tau(E)^2};\eqF \rfr{deltaerre} and \rfr{tra} tell us that at perihelion it amounts to \eqI\Delta(2\pi)=\Delta r(2\pi)\sqrt{1 + {4\pi^2}\rp{(1+e)}{(1-e)^3}}.\eqF
The angle $\xi$ between $\bds\Delta$ and $\bds r$ is given by
\eqI\tan\xi(E) = \rp{\Delta \tau(E)}{\Delta r(E)};\eqF at perihelion it is  \eqI\tan\xi(2\pi)=-2\pi\rp{\sqrt{1+e}}{(1-e)^{3/2}},\eqF i.e. $\xi$ is close to $-90$ deg; for the Earth it is $-81.1$ deg.
Thus, the difference $\delta $ between the lengths of the perturbed radius vector $r_{\rm pert}$ and the Keplerian one $r$ at a given instant amounts to about
\eqI \delta\approx\Delta\cos\xi;\eqF  in fact, this is precisely the quantity determined over 3 yr by the numerical integration of Figure \ref{deltaerref}. At the perihelion I have
\eqI \delta = \Delta r(2\pi)\sqrt{1 + {4\pi^2}\rp{(1+e)}{(1-e)^3}}\cos\xi;\eqF
since for the Earth \eqI\sqrt{1 + {4\pi^2}\rp{(1+e)}{(1-e)^3}}\cos\xi=1.0037,\eqF
it holds \eqI\delta\approx\Delta r(2\pi).\eqF This explains why Figure \ref{deltaerref} gives us just $\Delta r$.

Concerning the observationally determined increase of the Astronomical Unit, more recent estimates from processing of huge planetary data sets by Pitjeva  point towards a rate of the order of $10^{-2}$ m yr$^{-1}$ \cite{Pit05,Pit08}. It may be noted that my result for the secular variation of the terrestrial radial position on the line of the apsides would agree with such a figure by either assuming a mass loss by the Sun of just $-9\times 10^{-14}$ yr$^{-1}$ or a decrease of the Newtonian gravitational constant $\dot G/G\approx -1\times 10^{-13}$ yr$^{-1}$. Such a value for the temporal variation of $G$ is in agreement with recent upper limits from Lunar Laser Ranging \cite{LLR} $\dot G/G = (2\pm 7)\times 10^{-13}$ yr$^{-1}$.  This possibility is envisaged in \cite{dum} whose authors use
$\dot a/a =-\dot G/G$ by speaking about a small radial
drift of $-(6 \pm 13)\times 10^{-2}$ m yr$^{-1}$ in an orbit at 1 AU.
%
\section{The evolution of the Earth-Sun system}\lb{evol}
In this Section I will not consider other effects which may affect the final evolution of the Sun-Earth system like the tidal interaction between the
Earth and the tidal bulges of the giant solar photosphere, and the drag friction in the motion through the low chromosphere \cite{Sch08}.
For the Earth, by assuming the  values $a=1.00000011$ AU, $e=0.01671022$ at the epoch J2000 (JD 2451545.0) with respect to the mean ecliptic and
equinox of J2000 and $\dot\mu/\mu = -9\times 10^{-14}$ yr$^{-1}$, \rfr{mega} yields
\eqI\Delta r (2\pi) = 1.3\times 10^{-2}\ {\rm m}.\eqF   This means that at every revolution the position of the Earth is shifted along the true line
of the apsides (which coincides with the osculating one because of the absence of perihelion precession) by 1.3 cm.   This result is confirmed by my
numerical integrations and the discussion of Section \ref{minchia}; indeed, it can be directly inferred from Figure \ref{deltaerref} by multiplying
the value of $\Delta r$ at $t=1$ yr by $9\times 10^{-13}$.
By assuming that the Sun will continue to lose mass at the same rate for other 7.58 Gyr, when it will reach  the tip of the RGB in the HR diagram \cite{Sch08}, the Earth will be only $6.7\times 10^{-4}$ AU more distant than now from the Sun at the perihelion. Note that the value $9\times
10^{-14}$ yr$^{-1}$ is an upper bound on the magnitude of the Sun's mass loss rate; it might be also smaller \cite{Sch08} like, e.g., $7\times
10^{-14}$ yr$^{-1}$ which would yield an increment of $5.5\times 10^{-4}$ AU. Concerning the effect of the other planets during such a long-lasting
phase, a detailed calculation of their impact is beyond the scope of the present paper. By the way, I wish to note that the dependence of $\Delta
r(2\pi)$ on the eccentricity is rather weak; indeed, it turns out that, according to \rfr{mega}, the shift of the perihelion position after one orbit
varies in the range $1.3-1.1$ cm for $0 \leq e \leq 0.1$. Should the interaction with the other planets increase notably the eccentricity, the
expansion of the orbit would be even smaller; indeed, for higher values of $e$ like, e.g., $e=0.8$ it reduces to about 3 mm. By the way, it seems that
the eccentricity of the Earth can get as large as just $0.02-0.1$ \cite{Lask94,ito,Lask08} over timescales of $\approx 5$ Gyr due to the N$-$body
interactions with the other planets. In Table \ref{plane1}
I quote the expansion of the orbits of the other planets of the solar system as well.
\begin{table}
\caption{Expansion of the orbits, in AU, of the eight planets of the solar system in the next 7.58 Gyr for $\dot M/M=-9\times 10^{-14}$ yr$^{-1}$. I
have neglected mutual N-body interactions.}\lb{plane1}
\centering
\begin{tabular}{|l|l|}
\hline
\multicolumn{1}{|c|}{Planet} & \multicolumn{1}{c|}{$\Delta r$ (AU)} \\
\cline{1-2}
Mercury    & $2\times 10^{-4}$\\
Venus    & $5\times 10^{-4}$\\
Earth    & $7\times 10^{-4}$\\
Mars    & $9\times 10^{-4}$\\
Jupiter    & $3\times 10^{-3}$\\
Saturn    & $6\times 10^{-3}$\\
Uranus    & $1\times 10^{-2}$\\
Neptune    & $2\times 10^{-2}$ \\
 \hline
 \end{tabular}
\label{tavola1}
\end{table}
It is interesting to note that Mercury\footnote{It might also escape from the solar system or collide with Venus over $3.5$ Gyr from now \cite{Lask94,ito,Lask08}.} and likely Venus are fated at the beginning of the RGB; indeed, from Figure 2 of \cite{Sch08} it turns out that the Sun's photosphere will reach about $0.5-0.6$ AU, while the first two planets of the solar system will basically remain at 0.38 AU and 0.72 AU, respectively,
being the expansion of their orbits negligible according to Table \ref{plane1}.
After entering the RG phase things will dramatically change because in only $\approx 1$ Myr the Sun will reach the tip of the RGB phase loosing mass
at a rate of about $-2\times 10^{-7}$ yr$^{-1}$ and expanding up to 1.20 AU \cite{Sch08}. In the meantime, according to my perturbative
calculations, the perihelion distance of the Earth will increase by 0.25 AU. I have used as initial conditions for $\mu$, $a$ and $e$ their final
values of the preceding phase 7.58 Gyr-long.  In Table \ref{plane2} I quote the expansion experienced by the other planets as well; it is interesting
to note that the outer planets of the solar system will undergo a considerable increase in the size of their orbits, up to $7.5$ AU for Neptune,
contrary to the conclusions of the numerical computations in \cite{Dun} who included the mass loss as well. I have used as initial conditions the
final ones of the previous MS phase. Such an assumption seems reasonable  for the giant planets since their eccentricities should be left
substantially unchanged by the mutual N-body interactions during the next 5 Gyr and more \cite{Lask94,ito,Lask08}; concerning the Earth, should its
eccentricity become as large as $0.1$ due to the N-body perturbations \cite{Lask94,ito,Lask08}, after about 1 Myr its radial shift would be smaller
amounting to $0.22$ AU.
Mutual N-body interactions have not been considered.
thus hardly preventing our planet to escape from engulfment in the expanding solar photosphere.
Concerning the result for the Earth, it must be pointed out that it remains substantially unchanged if I repeat the calculation by assuming a
circularized orbit during the entire RGB phase.  Indeed, it is possible to show that by adopting as initial values of $a$ and $\mu$ the final ones of the
previous phase I get that after $\approx 1.5$ Myr $\Delta r$ has changed by 0.30 AU. Note that my results are in contrast with those in
\cite{Sch08} whose authors obtain more comfortable values for the expansion of the Earth's orbit, assumed circular and not influenced by tidal and frictional
effects, ranging from 1.37 AU ($|\dot\mu/\mu|=7\times 10^{-14}$ yr$^{-1}$) to 1.50 AU ($|\dot\mu/\mu|=8\times 10^{-14}$ yr$^{-1}$) and 1.63 AU ($|\dot\mu/\mu|=9\times 10^{-14}$ yr$^{-1}$).
 However, it must be noted that such a conclusion relies upon a perturbative treatment of \rfr{accel} and by assuming that the mass loss rate is
  constant throughout the RGB until its tip; in fact, during such a Myr the term $(\dot\mu/\mu)\Delta t$ would get as large as $2\times 10^{-1}$.
%
%
%
%
%
\begin{table}
\caption{Expansion of the orbits, in AU, of the eight planets of the solar system in the first 1 Myr of the RGB for $\dot M/M=-2\times 10^{-7}$
yr$^{-1}$. I have neglected mutual N-body interactions and other phenomena like the effects of tidal bulges and chromospheric drag for the inner planets.}\lb{plane2}
\centering
\begin{tabular}{|l|l|}
\hline
\multicolumn{1}{|c|}{Planet} & \multicolumn{1}{c|}{$\Delta r$ (AU)} \\
\cline{1-2}
Mercury    & $7\times 10^{-2}$\\
Venus    & $1.8\times 10^{-1}$\\
Earth    & $2.5\times 10^{-1}$\\
Mars    & $3.4\times 10^{-1}$\\
Jupiter    & $1.24$\\
Saturn    & $2.25$\\
Uranus    & $4.57$\\
Neptune    & $7.46$ \\
 \hline
 \end{tabular}
\label{tavola1}
\end{table}
In fact, by inspecting Figure 4 of   \cite{Sch08} it appears that in the last Myr of the RGB a moderate variation of $\dot M/M$ occurs giving rise to
an acceleration of the order of $\ddot{M}/M\approx 10^{-13}$ yr$^{-2}$. Thus, a further quadratic term of the form
\eqI \left(\rp{\ddot{\mu}}{\mu}\right)\rp{(t-t_0)^2}{2}\lb{qua}\eqF  should be accounted for in the expansion of  \rfr{accel}. A perturbative
treatment yields adequate results for such a phase 1 Myr long since over  this time span \rfr{qua} would amount to $\approx 5\times 10^{-2}$. However,
there is no need for detailed calculations: indeed,  it can be easily noted that the radial shift after one revolution is
\eqI\Delta r(2\pi)\propto \left(\rp{\ddot{\mu}}{\mu}\right)\rp{a^4}{\mu}.\lb{ddot}\eqF
After about 1 Myr \rfr{ddot} yields a variation of the order of $10^{-9}$ AU, which is clearly negligible.
\section{Conclusions}\lb{quattro}
I started in the framework of the two-body Newtonian dynamics by using a radial perturbing acceleration linear in time and straightforwardly treated it with the standard Gaussian scheme. I found that  the osculating semimajor axis $a$, the eccentricity $e$ and the mean anomaly $\mathcal{M}$ secularly decrease while the argument of pericentre $\omega$ remains unchanged; also the longitude of the ascending node $\Omega$ and the inclination $i$ are not affected.
The radial distance from the central body, taken on the fixed line of the apsides, experiences a secular increase $\Delta r$. For the Earth, such an effect amounts to about 1.3 cm yr$^{-1}$. By numerically integrating the equations of motion in Cartesian coordinates I found that the real orbital path expands after every revolution, the line of the apsides does not change and the apsidal period is larger than the unperturbed Keplerian one. I have also  clarified that such results are not in contrast with those analytically obtained for the Keplerian orbital elements which, indeed, refer to the osculating ellipses approximating the true trajectory  at each instant.
%
%
I applied my results to the evolution of the Sun-Earth system in the distant future with particular care to the phase in which the Sun, moved to the
RGB of the HR, will expand up to 1.20 AU in order to see if the Earth will avoid to be engulfed by the expanded solar photosphere. My answer is
negative because, even considering a small acceleration in the process of the solar mass-loss, it turns out that at the end of such a dramatic phase
lasting about $1$ Myr the perihelion distance will have increased by only $\Delta r\approx 0.22-0.25$ AU, contrary to the estimates in \cite{Sch08}
whose authors argue an increment of about $0.37-0.63$ AU.
In the case of a circular orbit, the osculating semimajor axis remains unchanged, as confirmed by a
numerical integration of the equations of motion which also shows that the true orbital period increases and is larger than the unperturbed Keplerian
one which remains fixed.
Concerning the other planets, while Mercury will be completely engulfed already at the end of the MS, Venus might survive;
however, it should not escape from its fate in the initial phase of the RGB in which the outer planets will experience  increases in the size of their
orbits of the order of $1.2-7.5$ AU.

As a suggestion to other researchers, it would be very important to complement my analytical two-body calculation by performing  simultaneous long-term numerical integrations of the equations of motion of all the major bodies of the solar system by including a mass-loss term in the dynamical force models as well to see if the N-body interactions in presence of such an effect  may substantially change the picture outlined here. It would be important especially in the RGB phase in which the inner regions of the solar system should dramatically change.

\section*{Acknowledgments}
I thank Prof. K.V. Kholshevnikov, St.Petersburg State University, for useful comments and references.


\end{document}